\documentclass[vecphys]{svmult}		% vectors in bold face
\usepackage{makeidx}         % allows index generation
\usepackage{graphicx,amsmath}        % standard LaTeX graphics tool
\usepackage{epsfig}          % another way to include figures
\usepackage{multicol}        % used for the two-column index
\usepackage[bottom]{footmisc}% places footnotes at page bottom

\makeindex             % used for the subject index
\begin{document}
\def\rcgindex#1{\index{#1}}
\def\myidxeffect#1{{\bf\large #1}}
%%%%%%%%%%%%%%%%%%%%%%%%%%%%%%%%%%%%%%%%%%%%%%%%%%%%%%%%%%%%%%%%%%%%%

% Title 
\title*{Nonreciprocal wave propagation through open, discrete nonlinear
Schr\"odinger dimers}
\titlerunning{Nonreciprocal wave propagation through DNLS dimers}

\author{Stefano Lepri\inst{1}
\and
Giulio Casati \inst{2,3}
}
\institute{Consiglio Nazionale delle
Ricerche, Istituto dei Sistemi Complessi, 
via Madonna del piano 10, I-50019 Sesto Fiorentino, Italy
\texttt{stefano.lepri@isc.cnr.it}
\and
Center for Nonlinear and Complex Systems, Universit\`a degli Studi
dell'Insubria, Como, Italy
\and
Istituto Nazionale di Fisica Nucleare, Sezione di Milano, Milan, Italy 
}

\maketitle
\abstract
{
We consider asymmetric (nonreciprocal) wave transmission  through a layered
nonlinear, non mirror-symmetric system described by the one-dimensional Discrete
Nonlinear Schr\"odinger equation with spatially varying coefficients embedded in
an otherwise linear lattice.  Focusing on the simplest case of two nonlinear
sites (the dimer), we compute exact scattering solutions
such that waves with the same frequency and incident amplitude impinging from
left and right directions have different transmission coefficients. 
The stability of some particular solutions is addressed. We show that
oscillatory instability may lead to the formation of stable
extended states coexisting with a nonlinear defect mode oscillating
at a different frequency. Numerical simulations of  
wave packet  scattering are presented. 
Gaussian wave packets with the same amplitude 
arriving from opposite directions on the dimer are indeed trasmitted 
very differently. Moreover, asymmetric transmission is sensitively
dependent on the input parameters, akin to the case of chaotic 
scattering. 
}

\section{Introduction}

The possibility to control energy and/or mass flows using nonlinear  features of
physical systems is a fascinating issue both from the point of view  of basic
science as well as from the applied one. In the context of wave propagation trough
nonlinear media, the simplest form of control would be to devise a ``wave
diode" \rcgindex{\myidxeffect{N}!Wave diode} in which electromagnetic or elastic waves are transmitted differently
along two opposite propagation directions. 

In a linear, time-reversal symmetric system this possibility 
is forbidden by the reciprocity theorem. As stated by Lord 
Rayleigh in his treatise \textit{The theory of sound} \cite{Rayleigh}:
\rcgindex{\myidxeffect{N}!Reciprocity theorem}
 \begin{quotation}
  Let A and B be two points [...] between which are situated 
  obstacles of any kind. Than a sound originating at A is 
  perceived at B with the same intensity as that with which 
  an equal sound originating at B would be perceived at A.
  In acoustics [...] in consequence of the not insignificant value of the 
  wavelength in comparison with the dimension of ordinary 
  obstacles the reciprocal relation is of considerable interest.
  \end{quotation} 

To achieve the desired effect  
one must thus violate the hypotesis of the 
theorem. In linear systems, a standard way is to break the time-reversal 
symmetry by applying a magnetic field
as done, for instance, in the case of optical isolators. 
An entirely alternative possibility is instead to consider \textit{nonlinear}
media. At least in principle, this option would offer a whole new range of 
possibilities of propagation control based on intrinsic material properties 
rather than an external field.  

Asymmetric wave propagation \rcgindex{\myidxeffect{N}!Asymmetric wave
propagation} induced by nonlinearity arises in several different
domains.  Among the  first examples discussed in the literature is the 
asymmetric phonon transmission through a nonlinear interface  layer between two
very dissimilar crystals \cite{Kosevich1995}.  In the field of nonlinear optics
a relatively vast number of approaches  exist. A so-called all-optical diode has
been proposed first in Ref.~\cite{Scalora94,Tocci95} and later on realized
experimentally \cite{Gallo01}. There are also proposals to employ
left-handed metamaterials \cite{Feise05}, quasiperiodic systems \cite{Biancalana08}, 
coupled nonlinear cavities \cite{Grigoriev2011} or $\mathcal{PT-}$symmetric
waveguides \cite{Ramezani2010,D'Ambroise2012}. 
Extension to the quantum regime in
which few-photon states display a diode effect has been proposed \cite{Roy2010}.

In the realm of acoustics the possibility of realizing a diode has been
demonstrated for nonlinear phononic media \cite{Liang09,Liang2010}. Another
promising context is the propagation of acoustic pulses through granular systems.
Indeed, experimental studies demonstrated a change of solitary wave reflectivity
from the interface of two granular media \cite{Nesterenko05}. More recently,
demonstration of rectification of  mechanical energy at sonic frequencies  in a
one-dimensional array of particles has been also reported \cite{Boechler2011}.

Despite the variety of physical contexts, the basic underlying  rectification
mechanisms rely on nonlinear phenomena as, for instance,  second-harmonic
generation in photonic \cite{Konotop02} or phononic crystals \cite{Liang09}, or
bifurcations \cite{Boechler2011}. In those examples the rectification depends
on  whether some harmonic (or subharmonic) of the fundamental wave is
transmitted or not. 

A related question is the possibility that the transmitted  power at the
\textit{same frequency and incident amplitude} would be sensibly different in
the two opposite propagation directions. In this Chapter we address the above
problem with the Discrete Nonlinear Schr\"odinger (DNLS) equation
\cite{Eilbeck1985,Kevrekidis} with site-dependent coefficients. It has been
demonstrated \cite{Kosevich02} that DNLS equation can be a sensible 
approximation for the evolution of longitudinal Bloch waves in layered photonic
or phononic crystals. Variable coefficients describe different nonlinear
properties of each layer and the presence of defects.  In the realm of the
physics of cold atomic gases, the equation is an approximate semiclassical
description of bosons trapped in periodic optical lattices (see e.g. Ref.
\cite{Franzosi2011} and references therein for a recent survey). Beyond its
relevance in many different physical contexts, the DNLS equation has the big
advantage of being among the simplest dynamical systems  amenable to a complete
theoretical analysis. For our purpose,  it is particularly convenient as it
allows to solve the scattering  problem exactly without the complications of
having to deal with wave harmonics \cite{Lepri2011}.

In Sec.~\ref{sec:model} we outline the model and show some examples of 
asymmetric plane-wave solutions. The issue of their stability is 
briefly addressed in Sec.~\ref{sec:stab}. It is also shown that 
oscillatory instability \rcgindex{\myidxeffect{N}!Oscillatory instability}
may lead to the formation of stable
extended states coexisting with a nonlinear defect mode oscillating
at a different frequency. In Sec.~\ref{sec:packet} we report some numerical
simulation of wave packet's scattering and illustrate its dependence on 
initial intensity. Finally, a brief 
summary of the results is given in Sec.~\ref{sec:end}.

%The discrete nonlinear Schr\"odinger (DNLS) equations should be written as
%
%\begin{equation}\label{DNLS}
%i\frac{d}{dt} \phi_{l,m,n} = i \dot\phi_{l,m,n} 
%= - C \Delta \phi_{l,m,n} \pm |\phi_{l,m,n}|^2 \phi_{l,m,n} =0,
%\end{equation}
%
%where $\Delta$ is the discrete Laplacian and $C$ the coupling constant.
%For 2D please use label indexes $(m,n)$ and for 1D use $n$.

\section{The model}
\label{sec:model}
  
The DNLS equation with spatially varying coefficients,
defined on an infinite one-dimensional lattice is given by
\rcgindex{\myidxeffect{N}!Scattering problem for DNLS}
\begin{equation}
i \dot \phi_n = V_n\phi_n 
-\phi_{n+1} - \phi_{n-1} + \alpha_n |\phi_n|^2 \phi_n \qquad, 
\label{tdnls}
\end{equation}
where units have been chosen such as the coupling $C=1$. 

We will assume the usual scattering setup where $V_n$ and $\alpha_n$ 
are non vanishing only for $1\le n \le N $. The two semi infinite portions 
($n<1$, $n>N$) of the lattice, model two leads where the wave can 
propagate freely \cite{Hennig99}. Let us look for solutions of the associated 
stationary transmission problem $\phi_n(t)=u_n \exp(-i\mu t)$
%(Fig.~\ref{fig:1})
\begin{equation}
\mu u_n = V_nu_n 
-u_{n+1} - u_{n-1} + \alpha_n |u_n|^2 u_n \qquad 1\le n \le N 
\label{dnls}
\end{equation}
of the form 
\begin{equation}
u_n = 
\begin{cases}
R_0 e^{ikn} + R e^{-ikn} & n\le 1 \\
T e^{ikn}                & n \ge N
\end{cases}
\label{wave}
\end{equation}
where $\mu = -2 \cos k$ and $0\le k \le \pi$ for the wave coming
from the left direction; $R_0, R$ and $T$
are the incident, reflected and transmitted amplitudes respectively.
The  solution sought must be complex in order to carry 
a non vanishing current $J=2|T|^2\sin k$. 

To break the mirror symmetry with respect to the center of the 
nonlinear portion, one must choose at least 
one of the two sets of coefficients $V_n$, $\alpha_n$
such that $V_n \neq V_{N-n+1}$, $\alpha_n \neq\alpha_{N-n+1}$. 
Note that the transmission of the right-incoming wave with the 
same $R_0$ and $\mu$ is computed by 
solving the problem with $(V_n,\alpha_n)\longrightarrow 
(V_{N-n+1},\alpha_{N-n+1})$ (i.e. ``flipping  the sample").
In the following, we will adopt the convention to label with $-k$ the 
right-incoming solutions with wave number $k$.
Nonlinearity is essential as for $\alpha_n=0$ the transmission 
coefficient is the same for waves 
coming from the left or right side, independently on $V_n$
due to time-reversal invariance of the underlying equations 
of motion \cite{Lindquist01}.

%\begin{figure}
%\center
%\includegraphics[width=10cm]{trasm.eps}
%\caption{The scattering setup}
%\end{figure}

The standard way to solve the problem is to introduce the (backward)
transfer map \cite{Delyon86,Wan90,Li96,Hennig99}
\rcgindex{\myidxeffect{N}!Transfer map}
%\begin{eqnarray}
%&&u_{n-1} = -v_n + (V_n-\mu + \alpha_n|u_n|^2)u_n \nonumber\\
%&&v_{n-1} = u_n
%\label{map}
%\end{eqnarray}
\begin{equation}
u_{n-1} = -v_n + (V_n-\mu + \alpha_n|u_n|^2)u_n, \quad
v_{n-1} = u_n \qquad.
\label{map}
\end{equation}
Note that these are 
complex quantities therefore the map is nominally four-dimensional.
However, due to conservation of energy and norm, it can be 
reduced to a two-dimensional area-preserving 
map \cite{Delyon86,Wan90,Li96,Hennig99} with 
an additional control parameter (the conserved current $J$). 
The solutions are straightforwardly found by iterating (\ref{map}) 
from the initial point $u_N = T \exp(ikN)$,  $v_N = T \exp(ik(N+1))$
dictated by the boundary conditions of Eq.~(\ref{wave}). 
For fixed $T$ and $k$, the incident and reflected 
amplitudes are determined as 
\begin{equation}
R_0 = \frac{\exp(-ik)u_0-v_0}{\exp(-ik)-\exp(ik)}, \quad
R = \frac{\exp(ik)u_0-v_0}{\exp(ik)-\exp(-ik)}
\nonumber
\end{equation}
and the transmission coefficient is $t(k,|T|^2) = |T|^2/|R_0|^2$
Note that if $(u_0,v_0)=(u_N,v_N)$ (periodic point of the map)
then $t=1$.

%\section{The dimer case}

For very short chains (oligomers), $t$ can be computed analytically. For
instance for the dimer $N=2$: 
\begin{equation}
%|R_0|^2 = |T|^2 \left|\frac{1+(\nu-e^{ik})(e^{ik}-\delta)}
%{e^{ik}-e^{-ik}}\right|^2
t = \left|\frac{e^{ik}-e^{-ik}}{1+(\delta_2-e^{ik})(e^{ik}-\delta_1)}
\right|^2
\end{equation}
where 
\begin{eqnarray*}
\delta_1 &=&V_2 -\mu +\alpha_2T^2, \\   
\delta_2 &=&V_1-\mu+\alpha_1T^2[1-2\delta\cos k +\delta_1^2].
\end{eqnarray*}
For the trimer $N=3$
\begin{equation}
t=\left|  \frac{e^{ik}-e^{-ik}}{e^{ik}-\delta_1 + (e^{ik}-\delta_3)(1-\delta_2(\delta_1-e^{ik}) ) }  \right|^2
\end{equation}
where
\begin{eqnarray*}
\delta_3 &=& V_3-\mu+\alpha_3|T|^2\nonumber\\
\delta_2 &=& V_2-\mu+\alpha_2 |T|^2 |\delta_3-e^{ik}|^2 \nonumber\\
\delta_1 &=& V_1-\mu+\alpha_1 |T|^2|1+\delta_2(e^{ik}-\delta_3)|^2.
\end{eqnarray*}
The formulas apply for $k>0$ (left-incoming waves); the transmission for 
right-incoming waves is obtained by exchanging the subscripts $1$ and $2$. 

\begin{figure}[htp]
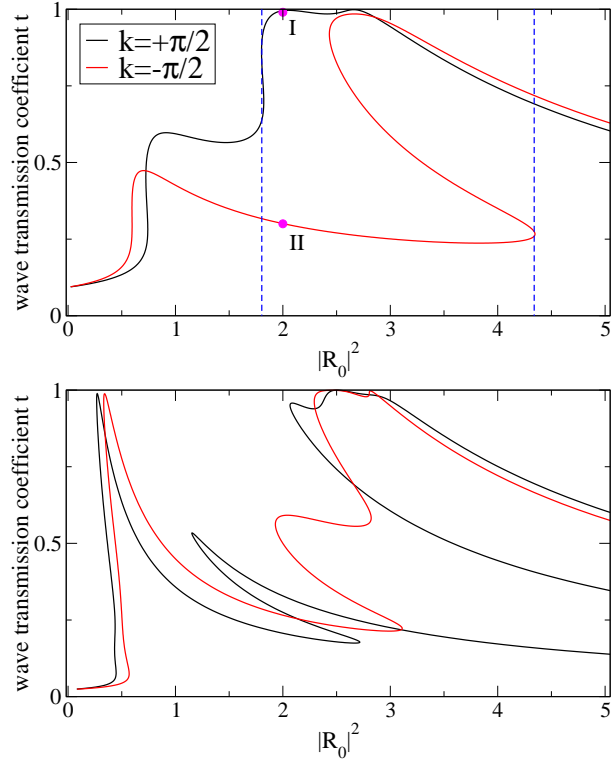

\begin{center}
 \includegraphics[width=8cm,clip]{Lepri_figure_01.eps}
 \includegraphics[width=8cm,clip]{Lepri_figure_02.eps} 
 \caption{Transmission coefficients as a function of the input intensity
 $|R_0|^2 $for $k=\pi/2$, $\alpha_n=1$. Upper panel: dimer
 $V_{1,2}=V_0(1\pm\varepsilon)$. Lower panel: trimer 
 $V_{1,3}=V_0(1\pm\varepsilon)$, $V_2=0$;
 $V_0=-2.5$ $\varepsilon=0.05$. }
 \label{fig:tco}
\end{center}
\end{figure}

Two examples of the dependence of $t$ on the input power are shown
in Fig.~\ref{fig:tco}. The curves display a multistable behavior
and, for strong enough intensities, are sizeably different indicating 
nonreciprocity. The effect is maximal in the vicinity of the 
nonlinear resonances that are detuned differently for the $k>0$ and $k<0$ 
cases yielding intervals of input values were multiple solutions
exist only for one propagation direction \cite{Lepri2011}.
\rcgindex{\myidxeffect{N}!Nonlinear resonances}
The lower panel of  Fig.~\ref{fig:tco} shows that even a moderate 
increase of the number of sites ($N=3$) dramatically increase the 
complexity of the curves as expected due to the mixed phase-space 
of the underlying transfer map \cite{Delyon86}.
\rcgindex{\myidxeffect{N}!Multistability}
%Since the phenomenon is of nonlinear origin the asymmetry depends
%on both frequency and amplitude. To quantify its efficiency, 
%in Fig.~\ref{f:fig2} we report the rectifying factor
%\begin{equation}
%f \;=\; \frac{t(k,|T|^2)-t(-k,|T|^2)}{t(k,|T|^2)+t(-k,|T|^2)}\quad,
%f \;=\; t(k,|T|^2)-t(-k,|T|^2)
%\label{rfactor}
%\end{equation}
%which approaches $\pm 1$ for maximal asymmetry. Note that,
%although increasing $\varepsilon$ broadens the regions 
%in which $|f|$ is relatively large, the overall transmitted intensity
%is reduced as well.   

To conclude this section, we mention that an alternative approach would
be to use the Green's function formalism previously used to compute the
stationary states for an electron moving on a chain with
nonlinear impurities \cite{Molina1993}. Indeed, for the case  of a
symmetric nonlinear dimer, resonance phenomena are demonstrated that lead to
complete transmission through the dimer \cite{Molina1993}. 
Of course, we expect such an approach to yield the same results
when applied to the present case.

\section{Stability of scattering solutions}
\label{sec:stab}

An important issue is the dynamical stability of the solutions.
To the best of our knowledge  no systematic study of scattering solutions of the 
type described above has been presented  in the literature (see
\cite{Malomed93} for the case of NLS equation with concentrated nonlinearities
continuum case and \cite{Miroshnichenko2009} for an analysis of a related 
problem, the nonlinear Fano effect).

The linear stability analysis is performed \cite{Eilbeck1985,Kevrekidis} 
by letting $\phi_n=(u_n + \chi_n)\exp(-i\mu t)$, and linearizing
the equation of motion to obtain 
\begin{equation}
i \dot \chi_n = (V_n-\mu)\chi_n 
-\chi_{n+1} - \chi_{n-1} + \alpha_n \left(2|u_n|^2 \chi_n 
+u_n^2 \chi_n^*\right)
\label{linear}
\end{equation}
Note that $\chi_n$ is complex. Letting 
\[
\chi_n \;=\; A_n \exp(i\lambda t) + B_n^* \exp(-i\lambda^* t)
\]
then $\lambda$ is solution of the eigenvalue problem
\begin{eqnarray}
&&\lambda A_n = -(\varepsilon_n - \mu) A_n + A_{n+1} + A_{n-1} -\theta_n B_n\nonumber\\
&&\lambda B_n = +(\varepsilon_n - \mu) B_n - B_{n+1} - B_{n-1} +\theta_n^* A_n
\label{eval}
\end{eqnarray}
with $\varepsilon_n\equiv V_n+2\alpha_n|u_n|^2$,  $\theta_n \equiv
\alpha_nu_n^2$. Note that, at variance with the  case of e.g. breather
solutions, the solutions are complex, and also the coefficients $\theta_n$ in
(\ref{eval}) are complex as well. As it is known, the eigenvalues come in
quadruplets of the form $\pm \lambda$, $\pm\lambda^*$. If eigenvalues have a
nonzero  imaginary part then the solution is unstable.  Generally speaking,
equilibria of an Hamiltonian system can lose  spectral (and therefore linear)
stability in two ways: a pair of real eigenvalues can  either (i) merge at the
origin and split onto  the imaginary axis (saddle-node bifurcation) or (ii)
collide at a nonzero  point and split off into the complex plane, forming a
complex  quadruplet (Krein bifurcation). The latter case correspond to
an oscillatory instability.

To solve the linear problem (\ref{eval}) exactly one should impose to
the solution a definite plane wave form (with complex wave numbers) in
the two seminiifinite linear parts of the chain. The matching of such waves
through the nonlinear portion reduces the infinite-dimensional problem
(\ref{eval}) to an homogeneous linear system of $2N$ equations, whose solvability
condition, along with the dispersion relations, yields a set of nonlinear
equations for the unknowns. The details of this method (which is technically
more involved than a straightforward diagonalization) will be presented
elsewhere \cite{future}. Here, we limit to illustrate the stability properties
of some  representative solutions by a more direct approach, i.e. by solving
numerically the eigenvalue problem for a finite truncation $-N_p\le n \le
N+N_p$  of the chain ($M$ sites with $M=2N_p+N+1$), checking that the relevant
eigenvalues of the resulting $2M\times 2M$ matrix are not affected by the
truncation error. 

\begin{figure}[htp]
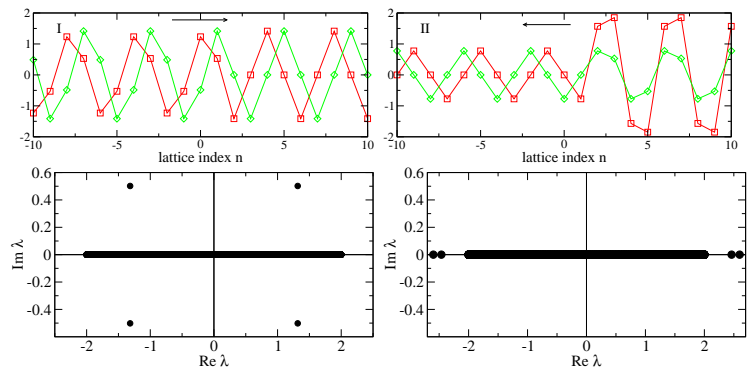

\begin{center}
 \includegraphics[width=0.4\textwidth,clip]{Lepri_figure_03.eps}
 \includegraphics[width=0.4\textwidth,clip]{Lepri_figure_04.eps}
 \includegraphics[width=0.41\textwidth,clip]{Lepri_figure_05.eps}
 \includegraphics[width=0.41\textwidth,clip]{Lepri_figure_06.eps}
 \caption{Upper panels: real (squares) and imaginary (diamonds) 
 parts of the two 
 solutions marked with 
 $I$ and $II$ in Fig.~\ref{fig:tco}: they correspond to incident
 waves with the same input $|R_0|^2=2$ having transmission
 coefficients $t=0.99$ and $t=0.30$ respectively.
 Lower panels: the spectrum in the complex plane 
 $N_p=200$. Isolated eigenvalues for for solution $I$ are $\pm 1.316  \pm 0.502i$
 for $II$, $\pm 2.60$ and $\pm 2.46$ respectively. 
  }
  \label{fig:sols}
\end{center}
\end{figure}

Of course, for small enough nonlinearities/amplitudes the solutions should be
stable. Since our main object of interest here is in the large 
asymmetry effects, we concentrate on the cases of strongly nonlinear
waves.  Fig. \ref{fig:sols} shows two examples of two such solutions
corresponding to the same  input (marked by dots in 
the upper panel of Fig.\ref{fig:tco}) along with their
eigenvalue spectra.  As expected, in both cases there is a continuum component
filling densely the  interval $[-2,2]$ on the real axis corresponding to
propagation of linear waves \footnote{Examining the eigenvalue spectra 
for different sizes $M$ of the matrix, reveals that this continuum
part of the spectra is mostly affected by truncation error. Typically the 
numerical eigenvalues have a spurious imaginary part which is of order $1/M$.
This is not surprising since the corresponding eigenvectors are extended 
waves and thus more sensitive to boundary effects.
}. In addition, isolated eigenvalues indicate
that the solution $I$ undergoes an oscillatory instability while $II$ is stable. 
The components  of the corresponding eigenvectors are exponentially localized
around the nonlinear portion of the chain. This is intuitively clear as only
perturbations located there can destabilize the solutions. 

\begin{figure}[htp]
\begin{center}
 \includegraphics[width=10cm,clip]{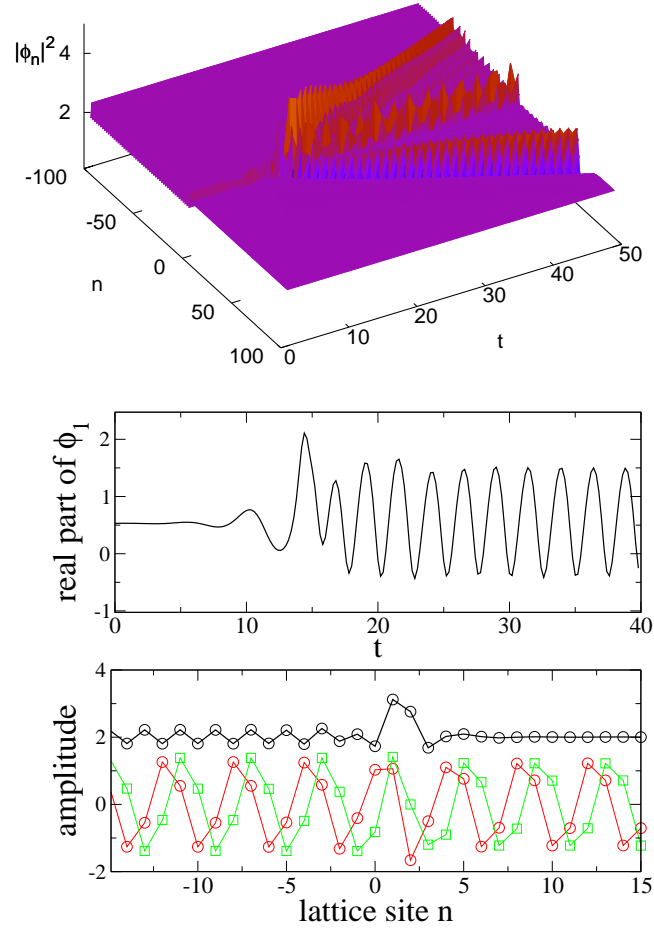}
 \includegraphics[width=8cm,clip]{Lepri_figure_08.eps}
 \includegraphics[width=8cm,clip]{Lepri_figure_09.eps}
 \caption{Upper panel: unstable evolution of the solution $I$ depicted in 
 Fig. \ref{fig:sols};
 numerical integration of DNLS with 
 initial condition corresponding to the stationary orbit
 with a small perturbation of size $10^{-5}$
 to the site $n=1$;
 Middle panel: evolution of 
 $Re \phi_1$: initial oscillations are at an angular frequency close 
 to the imaginary part of the unstable eigenvalue, $Re\lambda =1.316$. 
 At the later stage, a stable periodic oscillation sets in with 
 frequency increases to 2.5 which is outside 
 of the band of linear waves.
 Lower panel: snapshots of real (squares) and imaginary (diamonds) 
 parts of central part the chain at $t=400$. The upper curve
 is $|\phi_n|^2$ showing the appearance of a localized excitation
 residing on the dimer.  
 }
  \label{fig:inst}
\end{center}
\end{figure}

For comparison we also integrate the time-dependent DNLS equation setting as initial
condition $\phi_n(0)=u_n$ and imposing the boundary conditions 
$\phi_n(t) = u_n e^{-i\mu t}$ at the two edge sites
$n=-N_p$ and $n= N+N_p$ to simulate the infinite system.
As seen in Fig.\ref{fig:inst} the destabilization of solution $I$
occurs by an exponential growth of $\phi_n$ 
on the central sites accompanied by an 
oscillatory behavior with the frequency prescribed by the stability
spectrum (see the middle panel of Fig.\ref{fig:inst}). Since this frequency 
is in the band of linear waves, this process
is accompanied by emission of some radiation (traveling peaks in 
the upper panel of Fig.\ref{fig:inst}) until the amplitude 
and frequency become large enough leading to a stable localized 
object (lower panel in Fig.\ref{fig:inst}). This state 
is reminiscent of a nonlinear defect mode \cite{Kevrekidis}.  
There is however an important difference as the localized 
mode is superimposed to a plane wave 
and that the overall evolution is quasi-periodic.
\rcgindex{\myidxeffect{N}!Quasi-periodic solutions}
%From the point of view of the general motivation of our work, this observation
%is pretty intriguing as it shows that similar inputs can results also 
%in a more complex (quasi-periodic in this case) output.
%That is, the same wave can be modulated when impinging from one side but
%not from the other.

\section{Scattering of wave packets}
\label{sec:packet}

\begin{figure}[ht]
\includegraphics[width=12cm,clip]{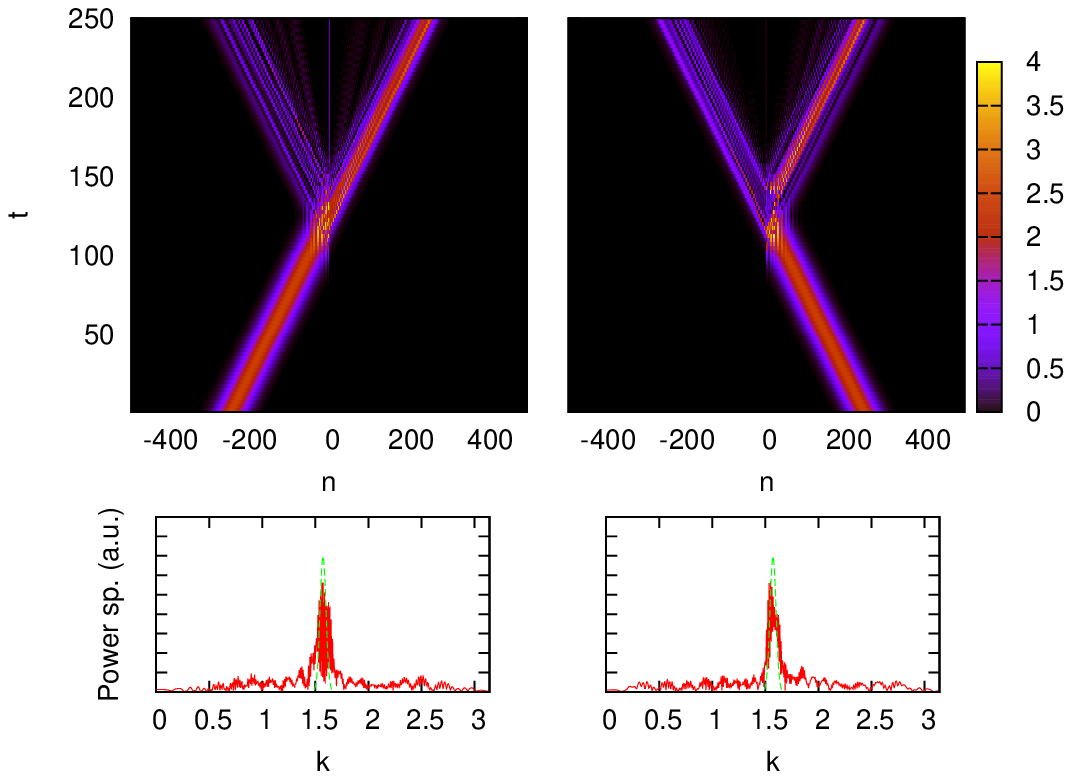}
\caption{Numerical simulations of the propagation 
of Gaussian wave packets, Eq.~(\ref{gauss}) 
impinging on a DNLS dimer. Here $V_0=-2.5$, $|k_0|=1.57$, 
$\varepsilon=0.05$, $M=500$, $|I|^2=2.5$, $w=50$ and 
$n_0=\mp 250$, respectively. 
Lower panels: Power spectra of the real part of $\phi_n$ at 
times $t=0$ (green dashed line) and $t=250$ (red solid line). 
}
\label{f:fig5}
\end{figure}

In this section we illustrate the consequences of the above results on the
transmission of wave packets. In a nonlinear system where the superposition
principle no longer holds,  the connection between the two problems is not
trivial. We solved numerically  the time-dependent DNLS  on a finite lattice
$|n|\le M$ with open boundary conditions, for the case of the dimer discussed in
\cite{Lepri2011}. We take as initial condition a Gaussian wave packet
(for $n_0<0$)
\begin{equation}
\phi_n(0)\;=\; I \exp \left[ -\frac{(n-n_0)^2}{w^2}+ik_0n \right]
\label{gauss}
\end{equation} 
where $k_0>0$ for $n_0 < 0$ (left-incoming packet) and 
where $k_0<0$ for $n_0 > N$ (right-incoming packet).
The upper panels of Fig.~\ref{f:fig5} display the evolution of two packets with
the same $I$ and opposite wave number $k_0$  impinging on the nonlinear dimer.
The asymmetry of  their propagation is manifest. In both cases, the packets are
significantly distorted after scattering, and the emerging envelope may 
vary wildly. However, the Fourier analysis shows that they
remain almost monochromatic at the incident wave number $k_0$ (lower panels of
Fig.~\ref{f:fig5}), with some small background amplitude radiation 
leaking throughout the lattice.
\rcgindex{\myidxeffect{N}!Wave packet scattering}

To quantify the asymmetry of the scattering, and to compare with the 
above analysis, we measured the 
wave packet transmission coefficient 
as the ratio between the transmitted norm 
at the end $t_{fin}$ of the run divided by the 
initial one, namely (for $n_0<0$)
\[
   t_{p}= \frac{\sum_{n>N}|\phi_n(t_{fin})|^2}{\sum_{n<1}|\phi_n(0)|^2}.
\]

\begin{figure}[htp]
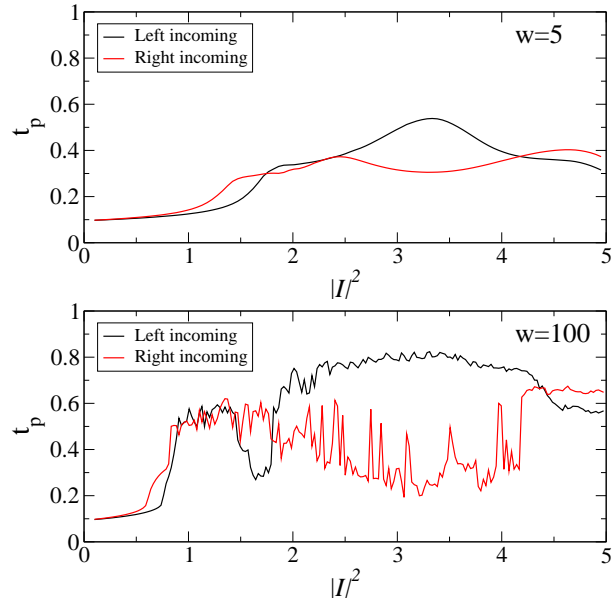

\begin{center}
 \includegraphics[width=8cm,clip ]{Lepri_figure_11.eps}
 \includegraphics[width=8cm,clip ]{Lepri_figure_12.eps}
 \caption{Dependence of wave packet transmission coefficients on
 intensity for two different wave packet widths $w$. Other parameters
 as in the previous figure.}
  \label{fig:tco2}
\end{center}
\end{figure}

In Fig. \ref{fig:tco2} we show the transmission coefficient as a function
of the packet intensity $|I|^2$ for two widths of the initial 
packet. Comparison with Fig.\ref{fig:tco} show that the region
of wave amplitudes
where the non-reciprocal behavior is maximal (that between the  vertical
dotted lines) corresponds qualitatively well to the range where packets 
with comparable intensities are transmitted more asymmetrically. 

Besides this, the wider packets  show a highly irregular behavior 
in the sense that there is \textit{sensitive dependence} on initial packets' parameters.  
This is somehow reminiscent of chaotic scattering (see e.g. \cite{Pikovsky1993}).
Empirically, we
found that this behavior depends very much on the width $w$: for small $w$ the curves are
much smoother.  Qualitatively speaking, this can be understood as follows.  The
dimer can be seen as an integrable, two-degrees of freedom system perturbed by a
time-dependent force (the incoming wave packet) and subject to dissipation (the
radiation towards the linear leads) \cite{Tietsche2008}. If the packet is very
narrow, the duration of the  perturbation is short and one may
argue that this just determines the initial condition for the dynamical system. 
The subsequent evolution will be almost regular and a slight change of the
forcing will have a little effect. On the contrary, a wider packet will result
in a more complex (possibly chaotic) dynamics yielding large changes in the
transmission.

\section{Summary and conclusions}
\label{sec:end}

We discussed the scattering problem for a linear Schr{\"o}dinger chain with an
embedded nonlinear, non-mirror symmetric dimer. The simplicity of the model
allows to compute analytically a whole family of non reciprocal plane wave
solutions, such as left- and right-incoming waves, 
with the same incident amplitude $R_0$ and
frequency $\mu$, have sizeably different transmission 
coefficients $t$. 

We than addressed the question of dynamical stability for some specific values
of $R_0$ and $\mu$. Solutions with a large enough $t$  generically undergo
oscillatory instabilities (see again Fig. \ref{fig:inst}). At the initial stage,
the development of the instability occurs through growth and oscillation
localized around the dimer sites, accompanied by  emission of radiation towards
infinity. Eventually, a stable extended state emerges coexisting with a
nonlinear defect mode oscillating at a different frequency. 
This numerical finding suggests that the lattice may support exact stable 
quasiperiodic, nonreciprocal solutions.  

Numerical simulations demonstrated that equal Gaussian wave packets impinging
on the dimer from the two opposite directions have indeed very different
transmission coefficients $t_p$. The packet amplitudes $I$ for which such
non-reciprocal behavior is maximal qualitatively  correspond to the range in
which  extended waves with the same wavenumber are transmitted more
asymmetrically. Moreover, $t_p$ is sensitively dependent on the input
parameters, like  for instance the initial width $w$ and amplitude $I$  (see
Fig.\ref{fig:tco2}), a feature which is reminiscent of chaotic scattering.

\section*{Acknowledgments}

SL thanks P.G. Kevrekidis and J. D'Ambroise for a useful correspondence  
regarding the stability problem. This work is part of the Miur PRIN 2008
project \textit{Efficienza delle macchine termoelettriche: 
un approccio microscopico}. 

% BibTeX users please use
 \bibliographystyle{unsrt}
 \bibliography{diodo,heat}{}

\printindex
\end{document}